\documentclass[runningheads]{llncs}

\usepackage{fourier}
\usepackage{syntax,etoolbox}
\AtBeginEnvironment{grammar}{\scriptsize}
\newcommand{\BS}{\char`\\}

\usepackage{amsmath,amssymb,mathtools,stmaryrd,upgreek}

\usepackage{newfloat}

\DeclareFloatingEnvironment[
  fileext   = logr,
  listname  = {List of Grammars},
  name      = Grammar,
  placement = htp
]{Grammar}
\DeclareFloatingEnvironment[
  fileext   = logr,
  listname  = {List of Denotations},
  name      = {$\uppi$ Denotation},
  placement = htp
]{PiDen}

\usepackage{multicol}
\usepackage{graphicx}

\usepackage{listings,xcolor}
\usepackage[colorlinks=true,urlcolor=blue,linkcolor=blue,citecolor=blue]{hyperref}

\usepackage{bbm}

\definecolor{LtGray}{rgb}{0.95,0.95,0.95}

\lstdefinestyle{padrao}{%
basicstyle=\scriptsize,	
columns=fullflexible,
mathescape=true,
tabsize=2, linewidth=\textwidth, 
numbers=left,				
numberstyle=\tiny,			
stepnumber=1,				
numbersep=5pt,			
backgroundcolor=\color{LtGray},
showspaces=false,			
showstringspaces=false,		
showtabs=false,			
frame=lines,				
tabsize=2,					
captionpos=b,				
floatplacement={tbp},
breaklines=true,			
breakatwhitespace=false,		
escapeinside={\%*}{*)},		
numberbychapter=false
}

\lstdefinestyle{maudedef}{%
basicstyle=\scriptsize\sffamily,	
columns=fullflexible,
mathescape=true,
tabsize=2, linewidth=\textwidth, 
showspaces=false,			
showstringspaces=false,		
showtabs=false,			
tabsize=2,					
captionpos=b,				
floatplacement={tbp},
breaklines=true,			
breakatwhitespace=false,		
escapeinside={\%*}{*)}		
}

\lstdefinelanguage{maude}{
    alsoletter={\:},
    morecomment=[l]{---},
    morecomment=[l]{***},
    keywords={pr, protecting, inc, including, sort, sorts, subsort, subsorts, including, class, msg, msgs, endfm, fmod, is, mod, endm, omod, endom},
    keywords=[2]{eq,  mb, ceq, if, rl, crl, else, then, fi, nonexec, variant},
    keywords=[3]{ctor, assoc, comm, gather, id\:},
    keywords=[4]{op, ops, var, vars, strat},
    keywords=[5]{search, red, reduce, rew, rewrite}
 }
\lstnewenvironment{maude}[1][]{\lstset{language=maude,style=padrao,#1}}{}
\lstnewenvironment{maudedef}[1][]{\lstset{language=maude,style=maudedef,#1}}{}

\lstdefinelanguage{AMN}{
	keywords={MACHINE, VARIABLES, INVARIANT, INITIALISATION, OPERATIONS, PRE, IF, =, /\, :=, OR, THEN, ELSE, END, BEGIN, WHILE, DO, CONSTANTS, VALUES}
}
\lstnewenvironment{amn}[1][]{\lstset{language=AMN,style=padrao,#1}}{}

\begin{document}

\frontmatter                                    

\mainmatter                                 

\title{B Maude: A formal executable environment for Abstract Machine Notation Descriptions}
\titlerunning{B Maude}

\author{Christiano Braga\inst{1} and Narciso Mart\'i-Oliet\inst{2}}
\authorrunning{C. Braga and N. Mart\'i-Oliet}

\institute{Universidade Federal Fluminense \\
\email{cbraga@ic.uff.br}
\and 
Universidad Complutense de Madrid \\
\email{narciso@ucm.es}
}

\maketitle

\begin{abstract}
We propose B Maude, a prototype executable environment for the Abstract Machine Notation implemented in the Maude language. B Maude is formally defined and results from the implementation of the semantics of AMN as denotations in the $\uppi$ Framework, a realization of Mosses' Component-based Semantics and Plotkin's Interpreting Automata. B Maude endows the B method with execution by rewriting, symbolic search with narrowing and Linear Temporal Logic model checking of AMN descriptions. 
\end{abstract}

\section{Introduction}

The B method~\cite{b-book} is an outstanding formal method for safety critical systems, in particular, railway systems. Such systems have quite strong requirements such as real-time constraints and fault (in)tolerance due to the nature of their problem domain that may involve lives. Therefore, they are natural candidates for the application of formal methods for their specification and validation.       

One of the main components of the B method are machine descriptions in the Abstract Machine Notation (AMN). They describe how a software component must behave in a safety-critical system.
Such descriptions are suitable for an automata-based interpretation and therefore can be naturally specified as transition systems~\cite{Clarke:2000:MC:332656} and validated by techniques such as model checking~\cite{Clarke:2000:MC:332656}.

This paper proposes B Maude, a tool for the validation of AMN descriptions. Our approach applies formal semantics of programming languages techniques. It is comprised by a set of basic programming languages constructs, inspired on Peter Mosses' Component-based Semantics~\cite{Mosses:2009:CS:1596486.1596489} whose semantics are given in terms of a generalization of Gordon Plotkin's Interpreting Automata~\cite{plotkin}. We named the resulting formalism $\uppi$ Framework~\cite{2018arXiv180504650B}. The semantics of AMN statements is given in terms of $\uppi$ constructions, or $\uppi$ Lib as we call it. B Maude is the implementation of the $\uppi$ Framework and the transformation from AMN to $\uppi$ in the Rewriting Logic language Maude~\cite{maude/2007}. It allows for execution by rewriting, symbolic search with narrowing and Linear Temporal Logic model checking of AMN descriptions. The B Maude prototype can be downloaded from \url{https://github.com/ChristianoBraga/BMaude}. 

This paper is organized as follows. In Section~\ref{sec:preliminaries} we recall the syntax of AMN and the use of Maude as a formal meta-tool. Section~\ref{sec:modpi} outlines the $\uppi$ Framework. Section~\ref{sec:pi-den-for-amn} formalizes AMN in $\uppi$ by giving denotations of AMN statements as $\uppi$ Lib constructions. Section~\ref{sec:bmaude} talks about B Maude by showing the application of search and model checking to an AMN description and outlines their implementation in B Maude. Section~\ref{sec:related-work} discusses work related to the formal semantics of AMN. Section~\ref{sec:conclusion} wraps up this paper with a summary of its contents and points to future work.

\section{Preliminaries}\label{sec:preliminaries}
%

\subsection{Abstract Machine Notation grammar and example}\label{sec:amn-cfg-and-ex}

In B Maude, we organize the Abstract Machine Notation (AMN) syntax in two parts. The first one is a context-free grammar for AMN expressions and commands, called Generalized Substitution Language. The second part declares machine level statements such as machines, variables, constants, values and operations. The AMN grammar will be used in Section~\ref{sec:pi-den-for-amn} when we specify the semantics of AMN as denotations in the $\uppi$ Framework. 
Only an excerpt is covered, due to space constraints, just enough to describe the example in Listing~\ref{amn:mutex}.


The Generalized Substitution Language, which essentially defines \emph{expressions} and \emph{commands}, is comprised by:
\begin{itemize}
\item identifiers, denoted by non-terminal $\langle\mathit{GSLIdentifiers}\rangle$\footnote{Non-terminals $\langle\mathit{ID}\rangle$, $\langle\mathit{RAT}\rangle$ and $\langle\mathit{BOOL}\rangle$ are left unspecified but simply denote what their names imply.}, 
\item arithmetic expressions, denoted by non-terminal $\langle\mathit{GSLExpression}\rangle$, predicates, denoted by non-terminal $\langle\mathit{GSLPredicate}\rangle$, and 
\item substitutions, denoted by non-terminal $\langle\mathit{GSLSubstitution}\rangle$. 
\end{itemize}
The notation for expressions and predicates is quite standard. Substitutions are essentially commands in AMN. Command `\_\texttt{:=}\_' denotes an assignment (an infix operator where `\_' denotes the positions of its operands),  `\texttt{IF$\_$THEN$\_$END}' and `\texttt{IF\_THEN\_ELSE\_END}' denote conditionals, and `\texttt{WHILE\_DO}' denotes unbounded repetition. 
Operator `\_\texttt{OR}\_' represents the bounded choice substitution. Keyword `\texttt{BEGIN\_END}' is a declaration and yields a scope where its substitutions must be evaluated within. The GSL grammar (excerpt) in BNF notation is as follows.

\begin{grammar}
<GSLIdentifiers> ::= `bid('<ID>`)' | `grat(' <RAT>`)' | `gboo(' <BOOL>`)' 

<GSLExpression> ::= <GSLIdentifiers> | <GSLPredicate> 
\alt <GSLExpression> `+' <GSLExpression> | $\ldots$

<GSLPredicate> ::= <GSLExpression>  `==' <GSLExpression> 
\alt <GSLExpression>  `/\BS'  <GSLExpression>  | $\ldots$

<GSLSubstitution> ::= 
<GSLIdentifiers> := <GSLExpression> 
\alt `IF'~<GSLPredicate>~`THEN'~<GSLSubstitution>~\\`ELSE'~<GSLSubstitution>~`END'
\alt `WHILE' <GSLPredicate> `DO' <GSLSubstitution>
\alt <GSLSubstitution> `OR' <GSLSubstitution> 
\alt `BEGIN' <GSLSubstitution> `END' <GSLSubstitution> | $\ldots$
\end{grammar}

The context-free grammar for AMN's machine-related statements
essentially defines \emph{declarations}:
\begin{itemize}
\item a `\texttt{MACHINE\_\_END}' declaration, denoted by non-terminal $\langle\mathit{AMNMachine}\rangle$, declares variables, constants, initializations for either or both, and operations,
\item a `\texttt{VARIABLES}' declaration (resp. `\texttt{CONSTANTS}'), in $\langle\mathit{AMNAbsVariables}\rangle$, is only a list of identifiers, and
\item a  $\langle\mathit{AMNValuation}\rangle$ declaration associates a substitution to the initialization of a variable or constant. An operation declaration, denoted by non-terminal $\langle\mathit{AMNOperation}\rangle$, associates a substitution to an (operation) identifier and a list of (identifier) parameters.
\end{itemize}
The AMN grammar in BNF notation is as follows.
\begin{grammar}
<AMNMachine> ::= `MACHINE' <GSLIdentifiers> <AMNClauses> `END'  | `MACHINE' <GSLIdentifiers> `END'  

<AMNClauses> ::=  <AMNAbsVariables> | <AMNAbsConstants> | <AMNOperations> | \\ <AMNValuesClause> | <AMNClauses> <AMNClauses>

<AMNAbsVariables> ::= `VARIABLES' <AMNIdList>

<AMNAbsConstants> ::= `CONSTANTS' <AMNIdList> 

<AMNIdList> ::= <AMNIdList> `,' <AMNIdList>

<AMNValuesClause> ::= `VALUES' <AMNValSet>

<AMNValuation> ::= <GSLIdentifiers> `=' <GSLExpression>

<AMNValSet> ::= <AMNValSet> `;' <AMNValSet>

<AMNOperations> ::= `OPERATIONS' <AMNOpSet>

<AMNOperation> ::= <GSLIdentifiers> `='  <GSLSubstitution>

<AMNOpSet> ::= <AMNOpSet> `;' <AMNOpSet>

<AMNOperation> ::= <GSLIdentifiers> `(' <AMNIdList> `)' <GSLSubstitution> 
\end{grammar}

\subsubsection{The \texttt{MUTEX} machine.}

As an illustrative example, Listing~\ref{amn:mutex} declares the \texttt{MUTEX} machine (which is actually executable in B Maude). It will be our running example in this paper. In Section~\ref{sec:bmaude}, we will use it to explain how B Maude is implemented. The \texttt{MUTEX} machine specifies a simple mutual exclusion protocol with two processes competing to enter a critical section. Each process, represented by an abstract variable in the \texttt{MUTEX} machine, can be in one of two possible states: \texttt{idle}, \texttt{wait}, or \texttt{crit}, encoded as constants in the machine. The protocol is encoded as the (parameterless) operation \texttt{mutex} that simply runs forever and, depending on the state of the processes, may non-deterministically change one of the processes state. For example, substitution 

{\scriptsize\begin{amn}
IF p1 == idle /\ p2 == idle THEN p1 := wait OR p2 := wait ELSE $\ldots$
\end{amn}}

\noindent declares that, in a situation where both \texttt{p1} and \texttt{p2} are in the \texttt{idle} state, either one or both of them may change to \texttt{wait}. 

\begin{amn}[caption=\texttt{MUTEX} protocol in the Abstract Machine Notation, label=amn:mutex]
MACHINE MUTEX
  VARIABLES p1 , p2
  CONSTANTS idle , wait , crit
  VALUES
    p1 = 0 ; p2 = 0 ;
    idle = 0 ; wait = 1 ; crit = 2
  OPERATIONS
    mutex =
      WHILE true DO
      BEGIN
				IF p1 == idle /\ p2 == idle 
				THEN (p1 := wait OR p2 := wait) 
				ELSE IF p1 == idle /\ p2 == wait 
				THEN p1 := wait OR p2 := crit
				ELSE IF p1 == idle /\ p2 == crit 
				THEN p1 := wait OR p2 := idle 
				ELSE IF p1 == wait /\ p2 == idle 
				THEN p1 := crit OR p2 := wait
				ELSE IF p1 == wait /\ p2 == wait 
				THEN p1 := crit OR p2 := crit 
				ELSE IF p1 == wait /\ p2 == crit 
				THEN p2 := idle 
				ELSE IF p1 == crit /\ p2 == idle 
				THEN p1 := idle OR p2 := wait 
				ELSE IF p1 == crit /\ p2 == wait 
				THEN p1 := idle 
				END END END END END END END END
			END
END
\end{amn}

\subsection{Maude as a meta-tool}\label{sec:pre-maude}

The Maude system and language~\cite{maude/2007} is a high-performance implementation of Rewriting Logic~\cite{meseguer92}, a formalism for the specification of concurrent systems that has been shown to be able to represent quite naturally many logical and semantic frameworks~\cite{MartiOliet2002}.


A (concurrent) system is specified by a rewrite system $\mathcal{R} = (\Sigma, E \cup A, R)$ where $\Sigma$ denotes its signature, $E$ the set of equations, $A$ a set of axioms, and $R$ the set of rules. The equational theory $(\Sigma, E \cup A)$ specifies the \emph{states} of the system, which are terms in the $\Sigma$-algebra modulo the set of $E$ equations and $A$ axioms, such as associativity, commutativity and identity. Combinations of such axioms give rise to different rewrite theories such that rewriting takes place \emph{modulo} such axioms. Rules $R$ specify the (possibly) non-terminating behavior, that takes place modulo the equational theory $(\Sigma, E \cup A)$. 
Computations in Maude are represented by rewrites according to either equations, rules or both in a given module. Functional modules may only declare equations while system modules may declare both equations and rules. Equations are assumed (that is, yield proof-obligations) to be Church-Rosser and terminating~\cite{Baader:1998:TR:280474}. Rules have to be coherent: no rewrite should be missed by alternating between the application of rules and equations. 

An equational theory $\mathcal{E} = (\Sigma, E \cup A)$ is declared in Maude as a \emph{functional} module with concrete syntax `\texttt{fmod  $\mathcal{E}$ is $I$ $\Sigma$ $E$ endfm}', where $I$ is the list of modules to be included by $\mathcal{E}$. A sort $s$ in $\Sigma$ is declared with syntax `\texttt{sort} $s$'. A subsort inclusion between $s$ and $s'$ is declared with syntax `\texttt{subsort $s$ < $s'$}'. An operation $o :\ \stackrel{\to}{s_i} \ \to s'$ in $\Sigma$ is declared with syntax `\texttt{op  $o$ : $\stackrel{\to}{s_i}$ -> $s'$ [$A^*$] .}', where $\stackrel{\to}{s_i}$ is the domain of $o$ denoted by the product of $s_i$ sorts, $1 \le i \le n$, $n \in \mathbbm{N}$, and $A^*$ is a list of $A$ attributes including `\texttt{assoc}', `\texttt{comm}', `\texttt{id: $c$}', `\texttt{idem}', that denote what their names imply, and $c$ is a constructor operator. Equations are declared with syntax `$\mathtt{eq} ~L~ \mathtt{=} ~R~.$' where $L$ and $R$ are terms in $\mathcal{T}_{\Sigma/A}(X)$, the $\Sigma$-algebra with variables in $X$, modulo axioms $A$. Equations can be conditional, declared, essentially, with syntax `\texttt{ceq $L$ = $R$ if $\bigwedge^n_i$ $L_i$ = $R_i$ .}', where $n \in \mathbbm{N}$, and $L_i$, $R_i \in \mathcal{T}_{\Sigma/A}(X)$. A rewrite theory $\mathcal{R} = (\Sigma, E \cup A, R)$ is declared in Maude as a \emph{system} module with concrete syntax `\texttt{mod  $\mathcal{R}$ is $I$ $\Sigma$ $E$ $R$ endm}' where $\Sigma$ and $E$ are declared with the same syntax as for functional modules and rules in $R$ are declared with syntax `\texttt{rl [l] : $L$ => $R$ .}' where $l$ is the rule label. As equations, rules can also be conditional, declared with syntax `\texttt{crl $L$ => $R$ if $\bigwedge^n_i$ $L_i$ = $R_i$ $\land$ $\bigwedge^m_j$ $L_j$ => $R_j$ .}', where $L_j$, $R_j \in \mathcal{T}_{\Sigma/A}(X)$ and $m \in \mathbbm{N}$.



One of the distinctive characteristics of Maude is the support to meta-programming. Meta-level applications in Maude, applications that use meta-programming,  apply the so called \emph{descent functions}~\cite[Ch.14]{maude/2007}. Such functions use a representation of modules as terms in a \emph{universal} theory, implemented in Maude as a system module called META-LEVEL in its prelude. Some of the descent functions are metaParse, metaReduce, metaRewrite and metaSearch. 
\begin{itemize}
\item Function metaParse receives a (meta-represented) module denoting a grammar, a set of quoted identifiers representing the (user) input and a quoted identifier representing the rule that should be applied to the given input qids, and returns a term in the signature of the given module. 
\item Descent function metaReduce receives a (meta-represented) module and a (meta-represented) term and returns the (meta-represented) canonical form of the given term by the exhaustive application of the (Church-Rosser and terminating) \emph{equations}, only, of the given module.  An interesting example of metaReduce is the invocation of the model checker at the meta-level: (i) first, module MODEL-CHECKER must be included in a module that also includes the Maude description of the system to be analyzed, and (ii) one may invoke metaReduce of a meta-representation of a term that is a call to function modelCheck, with appropriate parameters, defined in module MODEL-CHECKER. 
\item Function metaRewrite simplifies, in a certain number of steps, a given term according to both equations and rules (assumed coherent, that is, no term is missed by the alternate application of equations and rules) of the given module. 
\item The descent function metaSearch looks for terms that match a given \emph{pattern}, from a given term, according to a choice of rewrite relation from $\Rightarrow^*$, $\Rightarrow^+$, $\Rightarrow^!$, denoting the reflexive-transitive closure of the rewrite relation, the transitive closure of the rewrite relation or the rewrite relation that produces only canonical forms.
\end{itemize}

\subsection{Writing a compiler in Maude}\label{sec:comp-in-maude} 

A compiler for programs in a (source) language $L$ into programs in a (target) language $L'$ can be written in Maude as a meta-level application $\mathcal{C}$. The main components of $\mathcal{C}$ are: (i) a context-free grammar for $L$, (ii) the abstract syntax of $L$, (iii) a parser for programs in $L$ that generates abstract syntax trees for $L$, (iv) a transformer from the abstract syntax of a program in $L$ into the abstract syntax of a program in $L'$, (v) a pretty-printer for the abstract syntax of programs in $L'$, and (vi) a command-line user-interface for interacting with the compiler . 

The context-free grammar of a language $G = (V, T, P, S)$, where $V$ is the set of variables or non-terminals, $T$ is the set of terminals, $P$ is the set of productions of the form $V \to \alpha$, with $\alpha \in (V \cup T)^*$, and $S \not\in (V \cup T)$ the initial symbol, is represented, in Maude, as a functional module $\mathcal{G} = (\Sigma, \emptyset \cup A)$, that is, $E = \emptyset$, where, essentially, non-terminals are captured as sorts in $\Sigma$, non-terminal relations are captured by subsort inclusion, also in $\Sigma$, and terminals are represented as operations with appropriate signature in $\Sigma$, possibly with properties, such as associativity, declared in $A$. 

The parser for $L$ is a meta-function in a functional module $\mathcal{P} = (\Sigma, E)$ in Maude that includes, at least, the (i) functional module denoting the grammar of $L$, (ii) the functional module denoting the abstract syntax of $L$ and (iii) the functional module META-LEVEL. The set $E$ of equations in $\mathcal{P}$ are defined by structural induction on the syntax of $L$ \emph{encoded as meta-terms in Maude}, that is, they are such that the left-hand side of an equation is a meta-term denoting a statement in $L$ and its right-hand side is a term in the initial algebra of the functional module denoting the abstract syntax of $L$.

A transformer from the AST of $L$ to the AST of $L'$ is a meta-function in a functional module $\mathcal{T} = (\Sigma, E)$ including, at least, the functional modules for the AST for $L$ and $L'$ and the META-LEVEL. Each equation in $E$ is such that: (i) its left-hand side is given by a term with variables in the initial algebra of the functional module representing the AST of $L$, and, similarly, (ii) its right-hand side denotes a term with variables in the initial algebra of the functional module representing the AST of $L'$. When $L'$ is Maude itself, that is, the compiler generates \emph{a rewrite theory representing the rewriting logic semantics of} $L$, the many tools available in Maude, such as the rewrite module axioms engine, narrowing and Linear Temporal Logic model checker, are ``lifted'' to programs in $L$ through its rewriting logic semantics.

A pretty-printer for the AST of $L'$ is a meta-function in a functional module $\mathcal{PP} = (\Sigma, E)$ such that the equations in $E$ produce a list of quoted identifiers from a term in the initial algebra of the functional module denoting the AST of $L'$.

\section{$\uppi$ Framework}\label{sec:modpi}

In~\cite{2018arXiv180504650B}, the first author outlines the $\uppi$ Framework, a simple semantic framework for compiler construction based on Peter Mosses' Component-based Semantics (CBS)~\cite{Mosses:2009:CS:1596486.1596489} and Gordon Plotkin's Interpreting Automata~\cite{plotkin}.  The idea is to implement a formal core language that can be reused in the implementation of different compilers. The framework has two components: (i) $\uppi$ Lib, a library of basic programming language constructs inspired by Mosses' CBS and based on previous work of the first author with others, and (ii) $\uppi$ automata, an automata-based formalism for the specification of the semantics of programming language constructs that generalizes Plotkin's Interpreting Automata. Each component is discussed next, in Sections~\ref{sec:uppi-lib} and \ref{sec:uppi-automata}, respectively.

\subsection{$\uppi$ Lib signature}\label{sec:uppi-lib}

$\uppi$ Lib is a subset of Constructive MSOS~\cite{Mosses:2004:FCF}, as implemented in~\cite[Ch. 6]{msc-chalub}.
%
The signature of $\uppi$ Lib is organized in five parts, and implemented in four different modules in Maude: (i) expressions, that include basic values (such as rational numbers and Boolean values), identifiers, arithmetic and Boolean operations, (ii) commands, statements that produce side effects to the memory store, (iii) declarations, which are statements that construct the constant environment, (iv) output and (v) abnormal termination.  

Grammar~\ref{grm:uppi-lib} declares the CFG for $\uppi$ Lib identifiers, arithmetic expressions and commands. Given their simplicity, they  should be self-explanatory. Due to space constraints, we limit our exposition to these two syntactic classes only.
%
%
\begin{Grammar}
\begin{grammar}\scriptsize
<Control> ::= <Exp> | `ADD' | `SUB' | `MUL' | `DIV'  

<Exp> ::= <Id> | <AExp> | <BExp> 

<Id> ::= `idn' <ID> 

<AExp> ::= <RAT> | <AOp> <AExp> <AExp> 

<AOp> ::= `add' | `sub' | `mul' | `div' 

<Control> ::= <Cmd> | `ASSIGN' | `LOOP' | `IF'

<Cmd> ::= `nop' | `choice' <Cmd> <Cmd> | \\
`assign' <Id> <Exp> | `loop' <BExp> <Cmd>
\end{grammar}
\caption{$\uppi$ Lib excerpt}
\label{grm:uppi-lib}
\end{Grammar}
\subsection{$\uppi$ automata}\label{sec:uppi-automata}

$\uppi$ automata are a generalization of Plotkin's Interpreting Automata as defined in~\cite{plotkin}. 
Let $\mathcal{L}$ be a programming language generated by a Context Free Grammar (CFG) $G = (V, T, P, S)$ defined in the standard way where $V$ is the finite set of variables (or non-terminals), $T$ is the set of terminals, $P \subseteq V \times (V \cup T)^*$ and $S \not\in V$ is the start symbol of $G$.
An interpreting automaton for $\mathcal{L}$ is a tuple $\mathcal{I} = (\Sigma, \Gamma, \rightarrow, \gamma_0, F)$ where $\Sigma = T$, $\Gamma$ is the set of configurations, $\rightarrow \subseteq \Gamma \times \Gamma$ is the transition relation, $\gamma_0 \in \Gamma$ is initial configuration, and $F$ the unitary set of final configurations with the single element $\langle \emptyset, \emptyset, \emptyset \rangle$. Configurations in \(\Gamma\) are triples of the general form
\(
\Gamma = \mathit{Value~Stack} \times \mathit{Memory} \times \mathit{Control~Stack},
\)
where $\mathit{Value~Stack} = (L(G))^*$ with $L(G)$ the language accepted by $G$,
the set \(\mathit{Memory}\) is a finite map \(\mathit{Var}
\to_{\mathit{fin}} \mathit{Storable}\) with $\mathit{Var} \in V$ and $\mathit{Storable} \subseteq T^*$, and the
$\mathit{Control~Stack} = (L(G) \cup \mathit{KW})^*$, where $\mathit{KW}$ is the set of keywords of $\mathcal{L}$. 
A computation in $\mathcal{I}$ is defined as $\rightarrow^*$, the reflexive-transitive closure of the transition relation.

$\uppi$ automata are interpreting automata
whose \emph{configurations are sets of semantic components} that include, at least, 
a $\mathit{Value~Stack}$, a $\mathit{Memory}$ and a $\mathit{Control~Stack}$.
Plotkin's stacks and memory in
Interpreting Automata (or environment and stores of Structural
Operational Semantics) are generalized to the concept of \emph{semantic
  component}, as proposed by Peter Mosses in the Modular
SOS~\cite{Mosses:2004:MSOS} approach to the formal semantics of
programming languages.

Formally, a $\uppi$ automaton is an interpreting automaton $\mathcal{I}$ where, given an abstract preorder \emph{Sem}, for semantic components, the configurations of $\mathcal{I}$ are $\Gamma = \uplus^n_{i=1} \mathit{Sem}$, with $n
\in \mathbbm{N}$, $\uplus$ denoting the disjoint union operation of
$n$ semantic components, with $\mathit{Value~Stack}$, $\mathit{Memory}$ and $\mathit{Control~Stack}$ subsets of $\mathit{Sem}$.
%

Let us look now at the $\uppi$ automaton for the fragments of $\uppi$ Lib CFG in Grammar~\ref{grm:uppi-lib}, for arithmetic expressions and commands.
%
The values in the $\mathit{Value~Stack}$ are
elements of the set 
$\mathbbm{B} \cup \mathbbm{R} \cup \langle\mathit{Id}\rangle \cup \langle\mathit{BExp}\rangle \cup \langle\mathit{Cmd}\rangle,$ 
where \(\mathbbm{B}\) is the
set of Boolean values, \(\mathbbm{R}\) is the set of rational numbers. The $\mathit{Control~Stack}\) is defined as the set 
$(\langle\mathit{Cmd}\rangle \cup \langle\mathit{BExp}\rangle \cup \langle\mathit{AExp}\rangle \cup \mathit{KW})^*$, with 
$\mathit{KW}= \{\mathtt{ADD}, \mathtt{SUB}, \mathtt{MUL}, \mathtt{DIV}, \mathtt{ASSIGN},\mathtt{IF}, \mathtt{LOOP}\}$.

Informally, the computations of a $\uppi$ automaton mimic the behavior of a calculator
in Lukasiewicz postfix notation, also known as reverse Polish notation. A typical computation of a $\uppi$ automaton
\emph{interprets} a statement $c(p_1, p_2, \ldots, p_n) \in L(G)$ on the top of $\mathit{Control~Stack}$ \(C\) of a
configuration \(\gamma = (S, M, C) \cup \gamma'\), where $\gamma' \in \Gamma$, by unfolding its subtrees $p_i \in L(G)$ and $c \in \mathit{KW}$ that are then pushed back into $C$, possibly updating the $\mathit{Value~Stack}$
\(S\) with intermediary results of the interpretation of the $c(p_1, p_2, \ldots, p_n)$, and the $\mathit{Memory}$, should $c(p_1, p_2, \ldots, p_n) \in L(\langle\mathit{Cmd}\rangle)$.

For the transition relation of $\mathcal{I}$, let us consider the rules for
arithmetic sum expressions, 
\begin{eqnarray}
\label{eq:val}\langle S, M, n~ C \rangle \cup \gamma & \Rightarrow & \langle n ~ S, M, C \rangle \cup \gamma \\
\label{eq:add1}\langle S, M, \mathtt{add}(e_1, e_2) ~ C \rangle \cup \gamma & \Rightarrow &
\label{eq:add2} \langle S, M, e_1 ~ e_2 ~ \mathtt{ADD}~ C \rangle \cup \gamma \\
\langle n ~ m ~ S, M, \mathtt{ADD}~ C \rangle \cup \gamma & \Rightarrow & \langle (n + m)~ S, M, C \rangle \cup \gamma
\end{eqnarray}
where \(e_i\) are meta-variables for arithmetic expressions, and \(n, m
\in \mathbbm{R}\).  Rule~\ref{eq:add1} specifies that when the
arithmetic expression \(\mathtt{add(e_1, e_2)}\) is on top of the control
stack \(C\), then operator \texttt{ADD} should be pushed to \(C\) and then expression's operands \(e_1\) and \(e_2\) will be recursively
evaluated, as a computation is the reflexive-transitive closure of
relation \(\rightarrow\), leading to a configuration with an element in $\mathbbm{B} \cup \mathbbm{R}$ left on
top of the value stack $S$, as specified by
Rule~\ref{eq:val}. Finally, when $\mathtt{ADD}$ is on top of the control
stack $C$, and there are two natural numbers on top of $S$, they are
popped, added and pushed back to the top of $S$.

The rules for unbounded loop have standard meaning, as specified by Rules~\ref{eq:loop-unfold} to~\ref{eq:false-loop}. When a \texttt{loop} instruction is on top of the control stack, (i) the loop is pushed to the top of the value stack, the keyword \texttt{LOOP} and the loop's predicate expression (denoted by variable $E$ in Rule~\ref{eq:loop-unfold}) are pushed to the top of the control stack, (ii) if the recursive evaluation of $E$ produces \emph{true} on top of the value stack, the loop instruction and the body of the loop instruction are pushed to the top of the control stack, and (iii) if the recursive evaluation of $E$ produces \emph{false} on top of the control stack, the program evaluation continues with the remainder of the control stack $C$.
\begin{eqnarray}
\label{eq:loop-unfold}
\langle S, \mathtt{loop(E, K)}~ C \rangle \cup \gamma & \Rightarrow &
\langle (\mathtt{loop(E, K)}~ S), E ~\mathtt{LOOP}~ C \rangle \cup \gamma \\
\label{eq:true-loop}	
\langle \mathit{true} ~\mathtt{loop(E, K)}~ S, \mathtt{LOOP}~ C \rangle \cup \gamma & \Rightarrow &
\langle S, K ~\mathtt{loop(E, K)}~ C \rangle \cup \gamma \\
\label{eq:false-loop}
\langle \mathit{false} ~\mathtt{loop(E, K)}~ S, \mathtt{LOOP}~ C \rangle \cup \gamma & \Rightarrow &
\langle S, C \rangle \cup \gamma
\end{eqnarray}

\subsection{Model checking $\uppi$ automata}\label{sec:mc-pia}

Model checking (e.g.~\cite{Clarke:2000:MC:332656}) is perhaps the most popular formal method for the validation of concurrent systems. The fact that it is an \emph{automata-based automated validation technique} makes it a nice candidate to join a simple framework for language construction that also aims at validation, such as the one used in this paper.


This section recalls the syntax and semantics for (a subset of) Linear Temporal Logic, one of the modal logics used in model checking, and discusses how to use this technique to validate $\uppi$ automata, only the necessary to follow Section~\ref{sec:amn-mc}.

The syntax of Linear Temporal Logic is given by the following grammar 
\[\begin{array}{c}\scriptsize
\phi ::= \top ~|~ \bot ~|~ p ~|~ \neg(\phi) ~|~ (\phi \land \phi) ~|~ (\phi \lor \phi) ~|~ (\phi \to \phi) ~|~ 
(\Diamond \phi) ~|~ (\Box \phi) 
\end{array}\]
\noindent where connectives $\Diamond$, $\Box$
are called \emph{temporal modalities}. They denote ``Future state'' and ``Globally (all future states)''. There is a precedence among them given by: first unary modalities, in the following order $\neg$, $\Diamond$ and $\Box$, then binary modalities, in the following order, $\land, \lor$ and $\to$. 

The standard models for Modal Logics (e.g.~\cite{goldblatt}) are Kripke structures, triples $\mathcal{K} = (W, R, L)$ where $W$ is a set of worlds, $R \subseteq W \times W$ is the world accessibility relation and $L : W \to 2^{\mathit{AP}}$ is the labeling function that associates to a world a set of atomic propositions that hold in the given world. Depending on the modalities (or operators in the logic) and the properties of $R$, different Modal Logics arise such as Linear Temporal Logic.
A \emph{path} in a Kripke structure $\mathcal{K}$ represents a possible (infinite) scenario (or computation) of a system in terms of its states. The path $\tau = s_1 \to s_2 \to \ldots$ is an example. A \emph{suffix} of $\tau$ denoted $\tau^i$ is a sequence of states starting in $i$-th state.
Let $\mathcal{K} = (W, R, L)$ be a Kripke structure and $\tau = s_1 \to \ldots$ a path in $\mathcal{K}$. Satisfaction of an LTL formula $\phi$ in a path $\tau$, denoted $\tau \models \varphi$ is defined as follows, 
\[\begin{array}{l}
\tau \models \top, \qquad \tau \not\models \bot, \qquad \tau \models p ~\mathit{iff}~ p \in L(s_1), \quad \tau \models \neg\phi ~\mathit{iff}~ \tau \not\models \phi, \\
\tau \models \phi_1 \land \phi_2 ~\mbox{iff}~ \tau \models \phi_1 ~\mbox{and}~ \tau \models \phi_2, \\
\tau \models \phi_1 \lor \phi_2 ~\mbox{iff}~ \tau \models \phi_1 ~\mbox{or}~ \tau \models \phi_2, \\
\tau \models \phi_1 \to \phi_2 ~\mbox{iff}~ \tau \models \phi_2 ~\mbox{whenever}~ \tau \models \phi_1, \\
\tau \models \Box \phi ~\mbox{iff}~ \mbox{for all}~ i \ge 1, \tau^i \models \phi, \\
\tau \models \Diamond \phi ~\mbox{iff}~ \mbox{there is some}~ i \ge 1, \tau^i \models \phi. \\
\end{array}\]

A $\uppi$ automata, when understood as a transition system, is also a \emph{frame}, that is, $\mathcal{F} = (W, R)$, where $W$ is the set of worlds and $R$ the accessibility relation. A Kripke structure is defined from a frame representing a $\uppi$ automata by declaring the labeling function with the following state proposition scheme:
\begin{eqnarray}
\label{eq:state-prop}\forall \sigma \in \mathit{Memory}, v \in \mathit{Index}(\sigma), r \in \mathit{Storable}, \nonumber\\
\langle \sigma, \ldots \rangle \models p_v(r) =_{\mathit{def}} (\sigma(v) = r),
\end{eqnarray}
meaning that for every variable $v$ in the index of the memory store component (which is a necessary semantic component) there exists a unary proposition $p_v$ that holds in every state where $v$ is bound to $p_v$'s parameter in the memory store.  A \emph{poetic license} is taken here and $\uppi$ automata, from now on, refers to the pair composed by a $\uppi$ automata and its state propositions.
As an illustrative specification, used in Section~\ref{sec:amn-mc}, the LTL formula $\Box \neg[p_1(\mathit{crit}) \land p_2(\mathit{crit})]$ specifies safety (``nothing bad happens''), in this case both $p_1$ and $p_2$ in the critical section, when $p_i$ are state proposition formulae denoting the states of two processes and $\mathit{crit}$ is a constant denoting that a given process is in the critical section, and formula $\Box[p_1(\mathit{try}) \to \Diamond (p_1(\mathit{crit}))]$ specifies liveness (``something good eventually happens''),  by stating that if a process, $p_1$ in this case, tries to enter the critical section it will eventually do so.

\section{$\uppi$ denotations for the Abstract Machine Notation}\label{sec:pi-den-for-amn}

This section formalizes the meaning of B's Abstract Machine Notation (AMN) in terms of $\uppi$ denotations. The subset of the AMN considered in this section is (almost) in bijection with $\uppi$ Lib, so the formalization is quite easy to follow. 


\newcommand{\PIDEN}[1]{\llbracket #1 \rrbracket_\uppi}

The $\uppi$ denotational semantics of AMN, denoted by function $\llbracket\cdot\rrbracket_\uppi$, simply maps AMN statements, that is, arithmetic expressions, predicates, substitutions, machine  declarations, variable declarations, constant declarations, initialization declarations, and operation declarations, into $\uppi$ denotations in a quite direct way.
\begin{PiDen}
Let \texttt{F}, \texttt{I}, \texttt{P}, \texttt{V} in $\langle\mathit{GSLIdentifiers}\rangle$, \texttt{R} in $\langle\mathit{RAT}\rangle$, \texttt{B} in $\langle\mathit{BOOL}\rangle$, and \texttt{E}, $\mathtt{E_{i}}$ in $\langle\mathit{GSLExpression}\rangle$, $\mathtt{P}$ in $\langle\mathit{GSLPredicate}\rangle$, $\mathtt{S}$, $\mathtt{S_i}$ in $\langle\mathit{GSLSubstitution}\rangle$,  $\mathtt{A}, \mathtt{A_i}$ in $\langle\mathit{GSLActuals}\rangle$, $\mathtt{FS}$ in $\langle\mathit{AMNIdList}\rangle$, $\mathtt{O}$ in $\langle\mathit{AMNOperation}\rangle$, $\mathtt{OS}$ in $\langle\mathit{AMNOpSet}\rangle$, $\mathtt{VS}$ in $\langle\mathit{AMNValSet}\rangle$, $\mathtt{VC}$ in $\langle\mathit{AMNValuesClause}\rangle$, $\mathtt{AV}$ in $\langle\mathit{AMNAbsVariables}\rangle$, in

{\scriptsize \begin{align}
%
%
\llbracket \mathtt{I} \rrbracket_\uppi & =  \mathit{id}(\mathtt{I}), \\
\llbracket \mathtt{grat}(\mathtt{R}) \rrbracket_\uppi & =   \mathit{rat}(\mathtt{R}), \\
\llbracket \mathtt{gboo}(\mathtt{B}) \rrbracket_\uppi & =   \mathit{boo}(\mathtt{B}), \\
\label{eq:sum}\llbracket \mathtt{E_{\mathtt{1}} ~\mbox{\tt+}~ E_{\mathtt{2}}}\rrbracket_\uppi  & =  
	\mathit{add(\llbracket \mathtt{E_{\mathtt{1}}}\rrbracket_\uppi, \llbracket \mathtt{E}_{\mathtt{2}} \rrbracket_\uppi)}, \\
%
%
\llbracket \mathtt{E_{\mathtt{1}} ~\mbox{\tt==}~ E_{\mathtt{2}}}\rrbracket_\uppi  & =  
	\mathit{eq(\llbracket \mathtt{E_{\mathtt{1}}}\rrbracket_\uppi, \llbracket \mathtt{E}_{\mathtt{2}} \rrbracket_\uppi)}, \\
%
%
\label{eq:and}\llbracket \mathtt{E_{\mathtt{1}} ~\mbox{\tt/\BS}~ E_{\mathtt{2}}} \rrbracket_\uppi  & =  
	\mathit{and(\llbracket \mathtt{E_{\mathtt{1}}}\rrbracket_\uppi, \llbracket \mathtt{E}_{\mathtt{2}} \rrbracket_\uppi)}, \\
%
%
%
%
%
\label{eq:assign}\llbracket \mathtt{I ~\mbox{\tt :=}~ E}\rrbracket_\uppi  & =  
	\mathit{assign(\llbracket \mathtt{I} \rrbracket_\uppi, \llbracket \mathtt{E} \rrbracket_\uppi)}, \\
%
%
%
%
\llbracket \mathtt{S_{\mathtt{1}} ~\mbox{\tt OR}~ S_{\mathtt{2}}} \rrbracket_\uppi  & =  
	\mathit{choice(\llbracket \mathtt{S_{\mathtt{1}}} \rrbracket_\uppi, \llbracket \mathtt{S_{\mathtt{2}}} \rrbracket_\uppi)} , \\
%
%
%
%
%
%
%
\llbracket \mathtt{\mbox{\tt IF}~ P ~\mbox{\tt THEN}~ S_{\mathtt{1}} ~\mbox{\tt ELSE}~ S_{\mathtt{2}}} \rrbracket_\uppi  & =  
	\mathit{if(\llbracket \mathtt{P} \rrbracket_\uppi, \llbracket \mathtt{S_{\mathtt{1}}} \rrbracket_\uppi, \llbracket \mathtt{S_{\mathtt{2}}} \rrbracket_\uppi)}, \\
%
%
\llbracket \mathtt{\mbox{\tt WHILE}~ P ~\mbox{\tt DO}~ S} \rrbracket_\uppi  & =  
	\mathit{loop(\llbracket \mathtt{P} \rrbracket_\uppi, \llbracket \mathtt{S} \rrbracket_\uppi)} \\
%
%
%
%
\PIDEN{\mbox{\tt MACHINE} ~\mathtt{I} ~\mathtt{AV}~ \mathtt{C} ~\mathtt{VC}~ \mathtt{OP}~ \mbox{\tt END}} & =  
\mathit{dec}(\PIDEN{\mathtt{AV}, \mathtt{VC}}, \mathit{dec}(\PIDEN{\mathtt{C}, \mathtt{VC}}, \PIDEN{\mathtt{OP}})), \\
\PIDEN{\mbox{\tt MACHINE} ~\mathtt{I} ~\mathtt{AV}~\mathtt{VC}~ \mathtt{OP}~ \mbox{\tt END}} & =  
 \mathit{dec}(\PIDEN{\mathtt{AV}, \mathtt{VC}}, \PIDEN{\mathtt{OP}}), \\
\PIDEN{\mbox{\tt MACHINE} ~\mathtt{I} ~ \mathtt{C} ~\mathtt{VC}~ \mathtt{OP}~ \mbox{\tt END}} & =  
 \mathit{dec}(\PIDEN{\mathtt{C}, \mathtt{VC}}, \PIDEN{\mathtt{OP}}), \\
\PIDEN{\mbox{\tt MACHINE} ~\mathtt{I} ~ \mathtt{OP}~ \mbox{\tt END}} & =  \PIDEN{\mathtt{OP}}, \\
%
%
\PIDEN{\mbox{\tt VARIABLES} ~\mathtt{IS}, \mbox{\tt VALUES} ~\mathtt{VS}} & =  \PIDEN{\mathtt{IS}, \mathtt{VS}} , \\ 
%
%
\PIDEN{\mathtt{I}, \mathtt{I} \mbox{\tt =} \mathtt{E}} & =  \mathit{ref}(\PIDEN{\mathtt{I}}, \PIDEN{\mathtt{E}}), \\
\PIDEN{\mathtt{I}, (\mathtt{I} \mbox{\tt =} \mathtt{E} \mbox{\tt ;} \mathtt{VS})} & =  \mathit{ref}(\PIDEN{\mathtt{I}}, \PIDEN{\mathtt{E}}), \\
\PIDEN{(\mathtt{I} \mbox{\tt ,} \mathtt{IS}), (\mathtt{I} \mbox{\tt =} \mathtt{E} \mbox{\tt ;} \mathtt{VS})} & =  \mathit{dec}(\mathit{ref}(\PIDEN{\mathtt{I}}, \PIDEN{\mathtt{E}}), \PIDEN{\mathtt{IS}, \mathtt{VS}}), \\
\PIDEN{\mathit{I}, (\mathtt{I'} \mbox{\tt =} \mathtt{E} \mbox{\tt ;} \mathtt{VS})} & =  \PIDEN{\mathtt{I}, \mathtt{VS}}, \mathit{if} \mathtt{I} \not=\mathtt{I'} \\
\PIDEN{(\mathtt{I} \mbox{\tt ,} \mathtt{IS}), (\mathtt{I'} \mbox{\tt =} \mathtt{E} \mbox{\tt ;} \mathtt{VS})} & =  \PIDEN{(\mathtt{I} \mbox{\tt ,} \mathtt{IS}), \mathtt{VS}}, \mathit{if} \mathtt{I} \not=\mathtt{I'} \\
%
%
\PIDEN{\mbox{\tt OPERATIONS} ~\mathtt{O}} & =  \PIDEN{\mathtt{O}}, \\ 
\PIDEN{\mbox{\tt OPERATIONS} ~\mathtt{OS}} & =  \PIDEN{\mathtt{OS}}, \\ 
\PIDEN{\mathtt{O} \mbox{\tt ;} \mathtt{OS}} & =  \mathit{dec}(\PIDEN{\mathtt{O}}, \PIDEN{\mathtt{OS}}), \\ 
\PIDEN{\mathtt{P} \mbox{\tt=} \mathtt{S}} & =  \mathit{prc}(\PIDEN{\mathtt{P}}, \mathit{blk}(\PIDEN{\mathtt{S}})), \\
\PIDEN{\mathtt{P} \mbox{\tt(} \mathtt{FS} \mbox{\tt)} \mbox{\tt=} \mathtt{S}} & =  \mathit{prc}(\PIDEN{\mathtt{P}}, \PIDEN{\mathtt{FS}}^\mathit{for}, \mathit{blk}(\PIDEN{\mathtt{S}})), \\ 
\PIDEN{\mathtt{F}}^\mathit{for} & =  \mathit{par}(\PIDEN{\mathtt{F}}^\mathit{for}), \\
\PIDEN{\mathtt{F}\mbox{\tt ,} \mathtt{FS}}^\mathit{for} & =  \mathit{for}(\PIDEN{\mathtt{F}}^\mathit{for}, \PIDEN{\mathtt{FS}}^\mathit{for}), \\
%
%
\llbracket \mathtt{I} \mbox{\tt(}\mbox{\tt)}  \rrbracket_\uppi & =  \mathit{cal}(\llbracket \mathtt{I}\rrbracket_\uppi), \\
\llbracket \mathtt{I} \mbox{\tt(}\mathtt{E} \mbox{\tt)} \rrbracket_\uppi & =  \mathit{cal}(\llbracket \mathtt{I} \rrbracket_\uppi, \llbracket \mathtt{E}\rrbracket_\uppi ), \\
\llbracket \mathtt{I} \mbox{\tt(}\mathtt{A} \mbox{\tt)}  \rrbracket_\uppi & =  \mathit{cal}(\llbracket \mathtt{I}\rrbracket_\uppi, \llbracket \mathtt{A}\rrbracket^\mathit{act}_\uppi ), \\
\llbracket \mathtt{E} \rrbracket^\mathit{act}_\uppi & = \llbracket \mathtt{E} \rrbracket_\uppi, \\
\llbracket \mathtt{E} \mbox{\tt,} \mathtt{A} \rrbracket^\mathit{act}_\uppi & =  \mathit{act}(\llbracket \mathtt{E}\rrbracket_\uppi, \llbracket \mathtt{A}\rrbracket^\mathit{act}_\uppi ).
%
%
%
%
%
\end{align}}
\caption{Abstract Machine Notation}
\label{piden:amn}
\end{PiDen}

GSL expressions, such as sum and conjunction, are mapped to $\uppi$ Lib expressions, \emph{add} and \emph{and} in Equations~\ref{eq:sum} and~\ref{eq:and}. GSL substitutions are mapped to commands, such as \texttt{`\_:=\_}  being denoted by \texttt{assign} in Equation~\ref{eq:assign}. 


A `\texttt{MACHINE\_\_END}' declaration is associated with a \emph{dec} construction in $\uppi$ Lib. Such a declaration is a composition of further declarations for variables and constants, initialized according with the expressions in the `\texttt{VALUES}' clause, and operations. Variable declarations give rise to \emph{ref} declarations, that associate a location in the memory with a given identifier in the environment. (Constants are denoted by \emph{cns} $\uppi$ Lib declarations.) An operation declaration gives rise to a \emph{prc} declaration that binds an abstraction (a list of formal parameters together with a block of commands) to an identifier in the environment. An operation call substitution is translated into a \emph{cal} $\uppi$ Lib command, parametrized by the expressions resulting from the translations of the GSLExpressions given as actual parameters in the given operation call. 

\section{B Maude}\label{sec:bmaude}

B Maude is a formal executable environment for the Abstract Machine Notation that implements, in the Maude language, the $\uppi$ denotational semantics described in Section~\ref{sec:pi-den-for-amn}. 
The main objective of B Maude is to endow the B method with the validation techniques available in the Maude system, both the buit-in ones, such as rewriting, narrowing, rewriting modulo SMT and model checking, and user-defined ones.

\begin{figure}[ht]\centering
\includegraphics[width=0.5\textwidth]{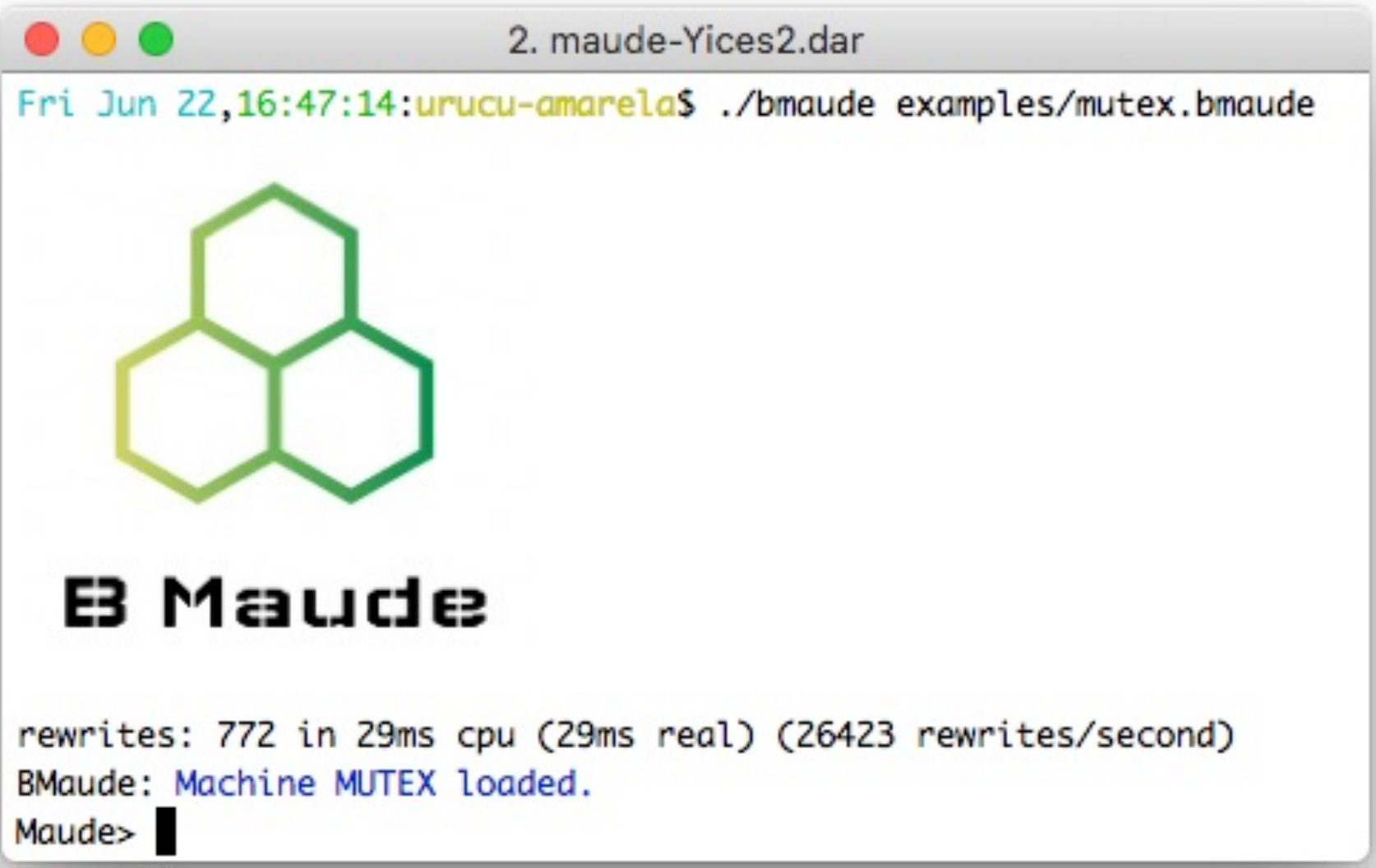}
\caption{B Maude banner and loading a machine}\label{fig:bmaude-splash-screen}
\end{figure}

The current version of \href{https://github.com/ChristianoBraga/BMaude}{B Maude} is called \textsc{uru\c{c}\'u amarela} (a bee common in the Rio de Janeiro area), it requires Maude Alpha 115 or later to run due to some of the narrowing-based techniques available in this version. (Even though it is not directly available from Maude's web site, Alpha 115 can be obtained free of charge by joining the Maude (Ab)users group by sending an email to {\small\email{maude-users@cs.uiuc.edu}}.) 
Figure~\ref{fig:bmaude-splash-screen} displays B Maude banner and the output of correctly loading the \texttt{MUTEX} machine. (The logo makes a ``visual pun'' with hexagons from a bee hive and components, the sound of the B method's name and its objective of giving support to component-based development.) 

B Maude's architecture is comprised essentially by the following components (recall the discussion in Section~\ref{sec:comp-in-maude}), together with the Maude implementation of the $\uppi$ Framework (see Section~\ref{sec:modpi}): (i) B Maude \emph{parser} that, given a list of (quoted) identifiers, produces a term in the initial algebra of the module \texttt{AMN-SYNTAX}, representing B Maude's grammar; 
(ii) B Maude to $\uppi$ Lib \emph{compiler}, that is exactly the implementation of the $\uppi$ denotations of Section~\ref{sec:pi-den-for-amn} in Maude; (iii) B Maude \emph{read-eval-loop}, a user interface that processes commands, given in terms of quoted identifiers, calls the appropriate meta-function; (iv) B Maude \emph{pretty-printer}, that translates the results coming from the meta-functions to user level representation, such as the output of the model checker in terms of $\uppi$ Lib constructions, into B Maude syntax. In the following sections we describe the implementation of some (due to space constraints) of the functionality implemented in the tool: 
(i) searching the state space, in Section~\ref{sec:search}, and (ii) model checking machines, in Section~\ref{sec:amn-mc}.

\subsection{Searching the computation graph of a GSL substitution} \label{sec:search}
\begin{figure}[ht]\centering
\includegraphics[width=.5\textwidth]{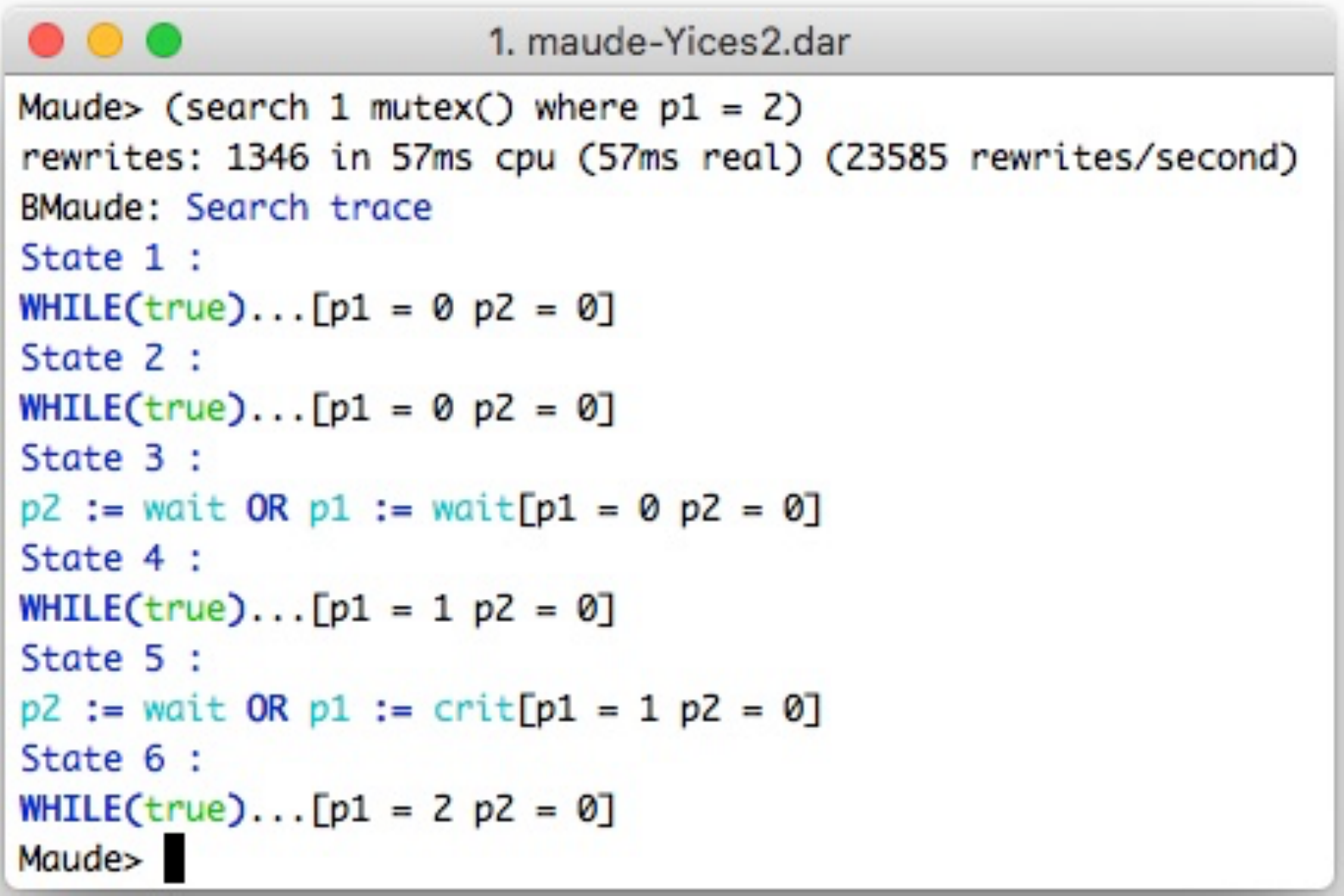}
\caption{Mutex simulation in B Maude}\label{fig:mutex-simulation}
\end{figure}

One may animate a B Maude description or check for invariants using the \texttt{search} command. Figure~\ref{fig:mutex-simulation} displays the execution of a \texttt{search} command querying for the first solution that satisfies the constraint \texttt{p1 = 2}, denoting the situation where process $1$ is in the critical section. 

The \texttt{search} command is handled in two levels. First, the syntax of the command is checked at module \texttt{BMAUDE-INTERFACE}, essentially to make sure that the conditions are well-formed. Then, if that is the case, rule \texttt{search}, in module \texttt{BMAUDE-COM\-MANDS}, is applied. First the descent function \texttt{metaSearch} is invoked by \texttt{bMaudeSearch}, with appropriate parameters. If \texttt{metaSearch} succeeds, then \texttt{bMaudePrintTrace} pret\-ty-prints the trace, from the \emph{initial state}, given by the $\uppi$ automaton state
\begin{maude}[caption=Initial state for \texttt{metaSearch}, label=lst:initial-state]
< cnt : blk(compile(M:AMNMachine), compile(S:GSLSubstitution)) ecs,
  env : noEnv, sto : noStore , val : evs ,
  locs : noLocs , out : evs , exc : CNT >,
\end{maude}
to the \texttt{N:Nat}-th solution, by invoking the descent function \texttt{metaSearchPath}. The initial state is such that the control stack \texttt{cnt} has a block resulting from the commands yielded by compilation (or $\uppi$ denotation) of the given GSL substitution \texttt{S:GSL\-Subs\-ti\-tu\-ti\-on} (a call to operation \texttt{mutex}, in Figure~\ref{fig:mutex-simulation}) together with the declarations resulting from the compilation of the machine \texttt{M:AMNMachine} (machine \texttt{MUTEX}, in this example) and remaining semantic components initialized to their default initial values. (For instance, \texttt{evs} is short for empty-value-stack and \texttt{CNT}, short for continue, means ``normal execution'', as opposed to an \texttt{EXT}, short for exit, meaning abnormal termination, that $\uppi$ Lib construction \emph{exit} may raise.)
\begin{maude}
crl [search] :
     < search N:Nat S:GSLSubstitution C:Condition ; M:AMNMachine ; QIL > =>
     < idle ; M:AMNMachine ;
             if T:ResultTriple? :: ResultTriple
             then
                bMaudePrintTrace(S:GSLSubstitution, M:AMNMachine, N:Nat) 	         
             else 
                printNoSolutionOrError(S:GSLSubstitution)
             fi >
if T:ResultTriple? := bMaudeSearch(S:GSLSubstitution, M:AMNMachine, N:Nat)
\end{maude}

\begin{figure}[ht]\centering
\includegraphics[width=.5\textwidth]{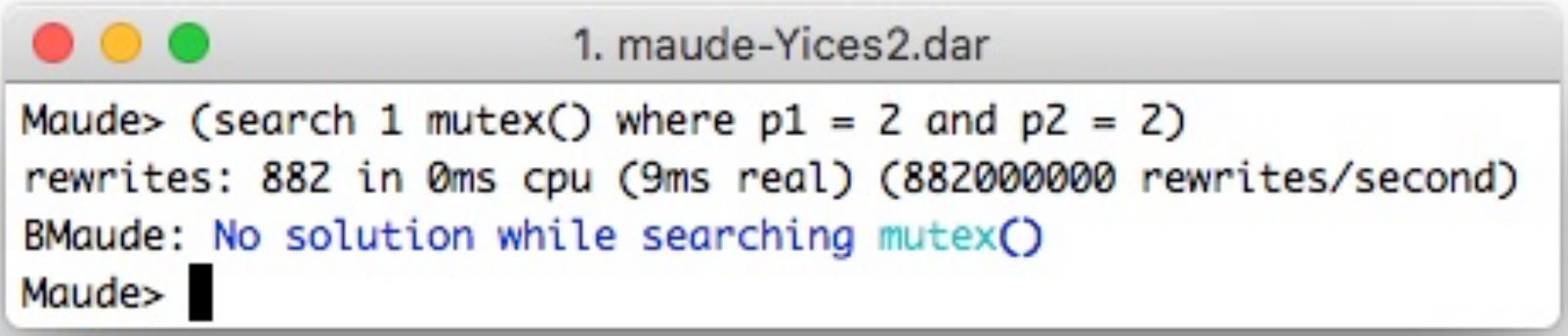}
\caption{Checking mutex safety invariant with search in B Maude}\label{fig:mutex-safety}
\end{figure}

Figure~\ref{fig:mutex-safety} is an example of the use of the \texttt{search} command to check for an \emph{invariant} property. In this example, a \emph{safety} property is checked by querying for a state such that both processes are in the critical section (that is, \texttt{p1 = 2 and p2 = 2}). Since no solution is found, then the protocol implemented by operation \texttt{mutex} is safe.
%

\subsection{Linear Temporal Logic model checking AMN Machines}\label{sec:amn-mc}

In Section~\ref{sec:mc-pia} we discussed how to model check $\uppi$ automata and in Section~\ref{sec:pi-den-for-amn} we gave denotations for Abstract Machine Notation (AMN) statements as $\uppi$ Lib constructions. Their combination yields the foundation to model check AMN descriptions for Linear Temporal Logic (LTL) properties. In B Maude, this is encoded as a meta-function that invokes the Maude LTL model checker to validate 
\[
\mathcal{K}(\llbracket \mathit{A} \rrbracket_\uppi), s_0 \models \varphi,
\]
where $\mathcal{K}(\llbracket \mathit{A} \rrbracket_\uppi)$ is the Kripke structure associated with the $\uppi$ denotations for AMN description $A$, $s_0$ is defined as in Listing~\ref{lst:initial-state}, and $\varphi$ is an LTL formula such that the atomic propositions are instances of the Equation Schema~\ref{eq:state-prop}, as defined in Section~\ref{sec:mc-pia}.

As with the \texttt{search} command, the model check command is handled in two levels. The first level makes sure that the command is well-formed. Then, Rule \texttt{mc} is applied and actually invokes the model checker by calling \texttt{metaReduce} with: (i) the module resulting from the application of operation \texttt{makeMCModule}, that creates a meta-module that  includes Maude's MODEL-CHECKER module and declares state propositions resulting from the variable declarations in $A$ and, (ii) a meta-term denoting an invocation of the \texttt{modelCheck} operation (in module MODEL-CHECKER) that model checks $\mathcal{K}(\llbracket \mathit{A} \rrbracket_\uppi)$, starting in the $\uppi$ state that has the $\uppi$ denotation of \texttt{S:GSLSubstitution} on top of the control stack, as in Listing~\ref{lst:initial-state}, for the LTL properties encoded in \texttt{T:Term}.
\begin{maude}[caption=]
crl [mc] :
     < mc S:GSLSubstitution T:Term ; M:AMNMachine ; QIL > =>
     < idle ; M:AMNMachine ;
             if (properResultPair?(T:ResultPair?)) 
             then printModelCheckResult(T:ResultPair?))	       
             else printModelCheckError(S:GSLSubstitution, T:Term) 
             fi >
if T:ResultPair? :=
   metaReduce(makeMCModule(M:AMNMachine),
     '_`,_|=?_[upTerm(M:AMNMachine), upTerm(S:GSLSubstitution), T:Term]) .
\end{maude}

\begin{figure}[ht]\centering
\includegraphics[width=.5\textwidth]{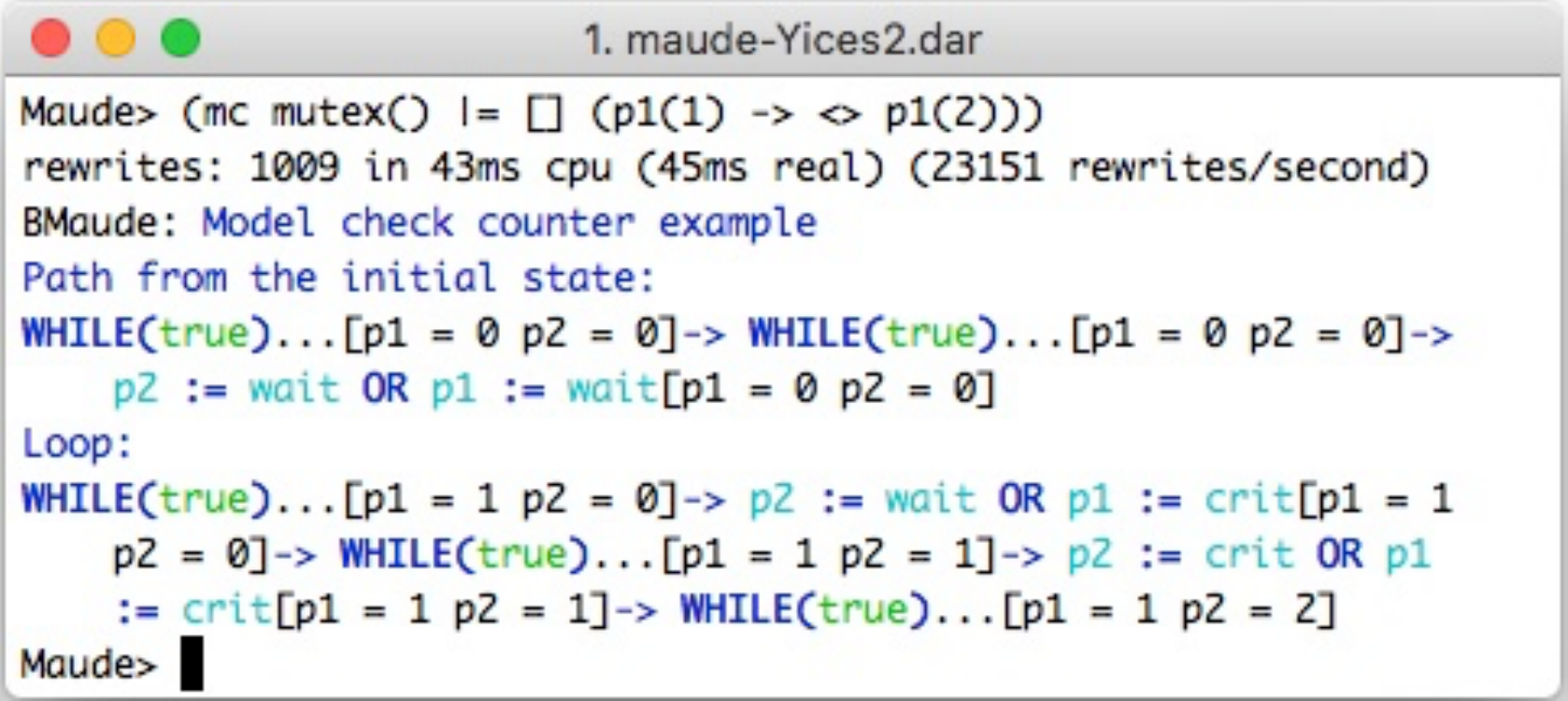}
\caption{Checking mutex liveness in B Maude}\label{fig:mutex-liveness}
\end{figure}

In Figure~\ref{fig:mutex-liveness} we see that machine \texttt{MUTEX} does not have the liveness property, that specifies that if a process tries to enter the critical section it will eventually do so, by showing a counter-example where the system enters a loop with only one of the processes, $\mathtt{p_2}$ in this scenario, accessing the critical section.

\section{Related work}\label{sec:related-work}

In~\cite{etmf:2016} the present authors together with David Deharbe and Anamaria Moreira proposed a Structural Operational Semantics for the Generalized Substitution Language (GSL). The SOS semantics for GSL had a straight forward representation in the Maude language by encoding SOS transition rules as conditional rules in Maude. Moreover, the Conditional Rewriting Logic Semantics of GSL in Maude was shown to have an equivalent Unconditional Rewriting Logic Semantics by considering conditional rewrites in the main rewrite ``thread''. The inspiration for the approach used there is the same here: Plotkin's Interpreting Automata. We have further developed this approach for operational semantics that is now called $\uppi$ Framework.
The semantics of the Abstract Machine Notation is then described as denotations in the $\uppi$ Framework, as described in Section~\ref{sec:pi-den-for-amn}.

GSL is presented in~\cite{b-book} with a weakest precondition (WP) semantics. Most of the work on B and GSL relies on this semantics. An interesting example is the work of Dunne in \cite{Dunne2002}, which includes additional information concerning the variables in scope at a substitution to solve some delicate issues that restrict what can be stated in B due to limitations of the pure WP semantics.  On the other hand, some related work proposing the embedding of B into other formalisms with strong tool support, such as Isabelle/HOL, can be found in the literature~\cite{Chartier1998,Deharbe2016}.  As in the current research, the purpose of those works is combining strengths of both worlds, to achieve further proof or animation goals. On a more theoretical line, but also with similar goals, the work in~\cite{Zeyda2005}  proposes a prospective value semantics to define the effect of GSL substitutions on values and expressions. The meaning of a computation, specified in GSL, is given in terms of the values of an expression, if the computation is carried out.  In the current paper we contribute to this line of work by discussing GSL operational semantics, its rewriting logic semantics, using Maude as the specification language for Rewriting Logic theories, and a prototype execution environment in the Maude system. 

\section{Conclusion}\label{sec:conclusion}

In this paper we have presented B Maude, an executable environment for the Abstract Machine Notation (AMN) language of the B method. Descriptions in AMN may be interpreted as automata and analyzed with automata-based methods such as model checking. We denote AMN statements as constructions in the $\uppi$ Framework, called $\uppi$ Lib constructions, which have an automata-based semantics given in terms of $\uppi$ automata. B Maude is then a compiler from AMN descriptions into $\uppi$ Lib constructions in Maude together with an implementation of the $\uppi$ Framework in Maude. B Maude allows for the execution by rewriting, symbolic search with narrowing and Linear Temporal Logic model checking of AMN descriptions by applying these techniques to the Maude representation of the given AMN description.

B Maude does not yet cover the complete AMN language and does not give support to the specification and reasoning on refinements. This is left to future work together with the inclusion of new validation techniques.

\paragraph*{Acknowledgements.} The $\uppi$ Framework is the result of a long term collaboration among the first author and Fabricio Chalub, Edward Hermann Haeusler, Jos\'e Meseguer and Peter D. Mosses, to whom he is deeply grateful.  



\def\sortunder#1{}

\end{document}